\documentclass[traditabstract]{aa}
\usepackage{graphicx}
\usepackage{txfonts}

 \newcommand{\mic}{$\mu$m}

\def\tex {\ifmmode{{T}_{\rm ex}}\else{$T_{\rm ex}$}\fi}
\def\tmb {\ifmmode{{T}_{\rm mb}}\else{$T_{\rm mb}$}\fi}
\def\ci     {\ifmmode{{\rm C}{\rm \small I}}\else{C\ts {\scriptsize I}}\fi}
\def\hi     {\ifmmode{{\rm H}{\rm \small I}}\else{H\ts {\scriptsize I}}\fi}
\def\hh     {\ifmmode{{\rm H}_2}\else{H$_2$}\fi}

\def\ts     {\thinspace}
\def\kms    {\ifmmode{{\rm \ts km\ts s}^{-1}}\else{\ts km\ts s$^{-1}$}\fi}
\def\msol   {\ifmmode{{\rm M}_{\odot}}\else{M$_{\odot}$}\fi}
\def\lsol   {\ifmmode{{\rm L}_{\odot}}\else{L$_{\odot}$}\fi}
\def\zsol   {\ifmmode{{\rm Z}_{\odot}}\else{Z$_{\odot}$}\fi}
\def\etal   {{\rm et\ts al.}}

\begin{document}

\title{Galaxy Evolution and Star Formation Efficiency at 0.2 $<$ z $<$ 0.6
\thanks{Based on observations carried out with the IRAM 30m telescope.
IRAM is supported by INSU/CNRS (France), MPG (Germany) and IGN (Spain)}}

\author{F. Combes \inst{1}
\and
S. Garc\'{\i}a-Burillo \inst{2}
\and
J. Braine \inst{3}
\and
E. Schinnerer \inst{4}
\and
F. Walter \inst{4}
\and
L. Colina  \inst{5}
           }
\offprints{F. Combes}
\institute{Observatoire de Paris, LERMA (CNRS:UMR8112), 61 Av. de l'Observatoire, F-75014, Paris, France
\email{francoise.combes@obspm.fr}
 \and
Observatorio Astron\'omico Nacional (OAN)-Observatorio de Madrid,
Alfonso XII, 3, 28014-Madrid, Spain 
 \and
Laboratoire d'Astrophysique de Bordeaux, UMR 5804,Universit\'e Bordeaux~I, BP 89, 33270 Floirac, France
 \and
Max-Planck-Institut f\"ur Astronomie (MPIA), K\"onigstuhl 17, 69117 Heidelberg, Germany
 \and
IEM, Consejo Superior de Investigaciones Cientificas (CSIC), Serrano 121, 28006 Madrid, Spain
              }

   \date{Received 10 September 2010/ Accepted 4 February 2011}

   \titlerunning{CO in ULIRGs at 0.2 $<$ z $<$ 0.6}
   \authorrunning{F. Combes et al.}

   \abstract{We present the results of a CO line survey
of 30 galaxies at moderate redshift (z $\sim$ 0.2-0.6), with the IRAM 30m
telescope, with the goal to follow galaxy evolution and in particular
the star formation efficiency (SFE) as defined by the ratio between
far-infrared luminosity and molecular gas mass (L$_{\rm FIR}$/M(\hh)).
 The sources are selected to be ultra-luminous infrared galaxies (ULIRGs),
with L$_{\rm FIR}$ larger than 2.8 10$^{12}$ \lsol,
experiencing starbursts;  adopting a low ULIRG CO-to-H$_2$ conversion
factor, their gas consumption time-scale is 
lower than 10$^8$ yr.  To date only very few CO observations exist
in this redshift range that spans nearly 25\% of the universe's age.
Considerable evolution of the star formation rate is already
observed  during this period.  18 galaxies out of our sample 
of 30 are detected (of which 16 are new detections), 
corresponding to a detection rate of 60\%.
The average CO luminosity for the 18 galaxies detected
is L'$_{\rm CO}$ = 2 10$^{10}$ K \kms\, pc$^2$, corresponding to an 
average \hh\, mass of  1.6 10$^{10}$ \msol.
 The FIR luminosity correlates well with the CO luminosity, in agreement with 
the correlation found for low and high redshift ULIRGs. 
Although the conversion factor between CO luminosity and \hh\, mass is uncertain,
we find that the maximum amount of gas
available for a single galaxy is quickly increasing as a function of redshift.
 Using the same conversion factor, the SFEs for z$\sim$0.2-0.6 ULIRGs are
found to be significantly higher, by a factor 3, than for local ULIRGs, and are comparable to 
high redshift ones. We compare this evolution to the expected cosmic
\hh\, abundance and the cosmic star formation history.

}
\maketitle


\section{Introduction}

 Ultra-Luminous Infra-Red Galaxies (ULIRGs) emit most of their energy
in the far-infrared, and have far-infrared luminosities 
L$_{\rm FIR}  > $  10$^{12}$ L$_\odot$,
(e.g. Sanders \& Mirabel 1996,  Veilleux \etal\, 2009). Since they can be seen so far away,
they allow us to explore the evolution of star
formation in the universe, and of the star formation efficiency (SFE)
in particular, defined as the FIR luminosity to \hh\, mass ratio (e.g. Kennicutt 1998).
Since the discovery of the first high redshift object in CO line emission
 (IRAS F10214+4724 at z=2.3, Brown \& Vanden Bout 1991, Solomon \etal\, 1992), 
there has been a wealth of CO-line discoveries,
a hundred objects are now detected at z$>$1, either from ULIRGs, or
from LIRGs (L$_{\rm FIR} >$ 10$^{11}$ L$_\odot$).
Some are amplified by gravitational lensing (see the
review by Solomon \& Vanden Bout 2005). They allow us to 
observe the interstellar medium of the galaxies, the CO excitation
(e.g. Weiss \etal\, 2007) and estimate
the amount of molecular gas present. Stars form from molecular gas,
so it is important to infer the \hh\, mass in order to determine the SFE.
At high redshift, many of these objects are quasars or radio-galaxies
(due to their selection, e.g. Omont \etal\, 2003), however, their
FIR emission is powered predominantly by star formation (e.g. 
Riechers \etal\, 2009, Wang \etal\, 2010).

Locally, our knowledge of the ULIRG phenomenon is more profound due to
higher spatial resolution and sensitivity. 
Because the peak of the 
dust emission is progressively shifted from the FIR to the submm domain,
the dust emission can be detected to high redshifts
(negative K-correction, e.g. Blain \& Longair 1996).
The CO-line emission is less favoured, and CO lines are 
difficult to detect at high z although observing
the high-J CO lines helps significantly in highly excited objects (Combes \etal\, 1999).
To date, more than a hundred objects have been studied in detail locally. In the case of
the ULIRGs, it was found that they are characterized by compact, nuclear starbursts
(e.g. Downes \& Solomon 1998), and it has been argued that a special
CO-to-\hh\, 
conversion factor should be used, that is 5.75 lower than the standard
factor commonly used for  Milky Way-like galaxies
(Downes \etal\, 1993). In the present paper, we
will adopt for ULIRGs the ratio $\alpha$=0.8 (Solomon \etal\, 1997)
between M(\hh) and L'$_{\rm CO}$, expressed in units of M$_\odot$ (K \kms\, pc$^2$)$^{-1}$,
and not the standard $\alpha$=4.6. 

At intermediate redshifts, between 0.2 $<$ z $<$ 1, there is a dearth of
CO-line detections. This is partly due to observational difficulties.
The most commonly used millimetric window is the 3mm one, which is least affected by 
atmospheric opacity. Between 81 and 115\,GHz, all redshifts can be observed with 
at least one line of the CO rotational ladder, except between z=0.4 and 1.
 The latter can be observed at 2mm (targeting the CO(2-1) line), but in less
favorable atmospheric conditions. While between z=0.2 and z=0.4
the 3mm window can be used in the CO(1-0) transition, the K-correction
is strongly reducing its observable intensity 
(by a factor growing faster than (1+z)$^4$, Combes \etal\, 1999).
The  redshift range 0.2 $<$ z $<$ 1 is important though, as 
 it covers almost half of the age of the universe, and also the most dramatic
 change in star formation activity (e.g. Madau \etal\, 1998, Hopkins \& Beacom 2006). 
In the universal star formation history, the most striking feature is
the impressive drop between z=1 and z=0
by at least an order of magnitude (Blain \etal\, 1999).

Up to now, very little was known about the molecular gas content of
galaxies at moderate redshift. The ULIRG sample of Solomon \etal\, (1997)
contains 37 objects, but only 2 have  $z > 0.2$. Negative results were obtained 
in previous studies, conducted about 10yrs ago (Lo \etal\, 1999, Wilson \& Combes 1998),
but the performances of the mm-instruments have dramatically improved since then.
Two more objects were detected by Geach \etal\, (2009), although more upper
limits were also reported (Melchior \& Combes 2008).
To study star forming galaxies in this period,
and in particular to derive their star formation efficiency, we have
undertaken a CO-line search in the range 
$0.2 < z < 0.6$,  almost unknown territory as far as molecular
lines are concerned. We have selected a sample of 30 
IR-luminous galaxies in this redshift range to check whether the
variation of star forming activity  is due to a variation in
molecular gas content or star formation efficiency, or both.
One of the objects (IRAS 11582+3020, hereafter G4) has already been mapped
with the IRAM Plateau de Bure Interferometer (Combes \etal\, 2006, paper I).
 The CO map showed spatially resolved emission on 30kpc scales
and revealed a velocity gradient.  It was concluded in that paper
that not all the molecular gas is confined in a nuclear starburst,
but that $\sim$50\% of it is extended on galactic scales (25-30kpc).
In the present paper, we describe the CO survey carried out with the IRAM
30m telescope. The sample is described in Sect. \ref{sample} and the observations
in  Sect. \ref{obs}. Results are presented in  Sect. \ref{res} and
discussed in Sect. \ref{disc}. 

\begin{figure}[htp]
\label{fig:sample}
\includegraphics[angle=-90,width=8cm]{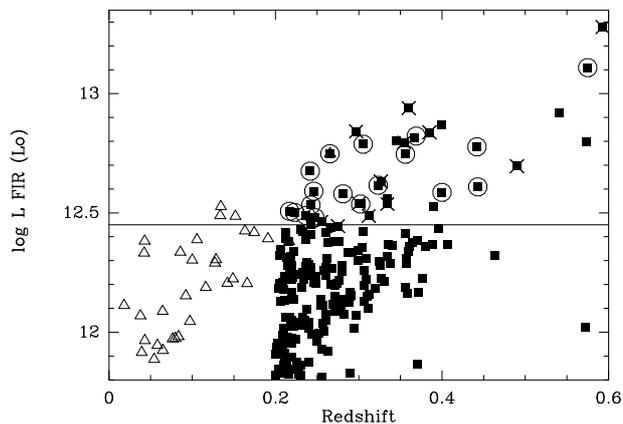}
\caption{Definition of our sample. Among the 209 northern galaxies (filled symbols)
found in NED between 0.2 $<$ z $<$ 0.6 and detected at 60 \mic\, by IRAS,
we selected the most luminous ones (log $L_{\rm FIR} /\lsol >$ 12.45, 
as indicated by the horizontal line).
By comparison, the ULIRGs in the sample of Solomon \etal\, (1997) are plotted as
open triangles. The circles indicate detections,
non-detections are marked by a cross. Sources that have neither a circle nor a cross could not
be observed due to weather conditions. }
\end{figure}

\section{The sample}
\label{sample}

  The present-day sensitivity in the CO line  restricted the sample to
the brightest objects in the far infrared.
We have selected all objects between  $0.2 < z <0.6$
 and DEC(2000) $>$ -12$^\circ$ that are identified as galaxies in the literature, have
 spectroscopic redshifts, and are detected at 60\mic\,  (IRAS, ISO).
This resulted in a total of 209 galaxies.
Most of the galaxies (and in particular the brightest ones) have
detailed photometry in the NIR bands, from the samples by
Clements \etal\, (1996), Kim \& Sanders (1998), Kim \etal\, (2002) 
and Stanford \etal\, (2000). 
The available sub-arcsec K-band images (from either IRTF or Keck telescopes,
in the above references),
reveal that about two-thirds of the objects are interacting galaxies. 

Out of the 209 galaxy sample, we selected the brightest ones,
with log L$_{\rm FIR} >$ 12.45. This leads 
to a sample of 36 objects, to be observed with the 30m telescope. 
We did not reobserve one galaxy detected by Solomon \etal\, (1997), 
nor 3C48, detected by Wink \etal\, (1997), although we include these 
2 sources in our analysis. Due to
weather conditions, only 28 sources were observed in the project. All
their identification and coordinates are displayed in
Table \ref{tab:sample}.
Out of the 30 objects in the sample (including the two literature sources), 18
were detected, corresponding to a detection rate of 60\%.
Figure  \ref{fig:sample} displays the distribution of FIR luminosities with redshift.

\begin{figure}[!t]
\resizebox{8cm}{!}{\includegraphics[angle=0,width=8cm]{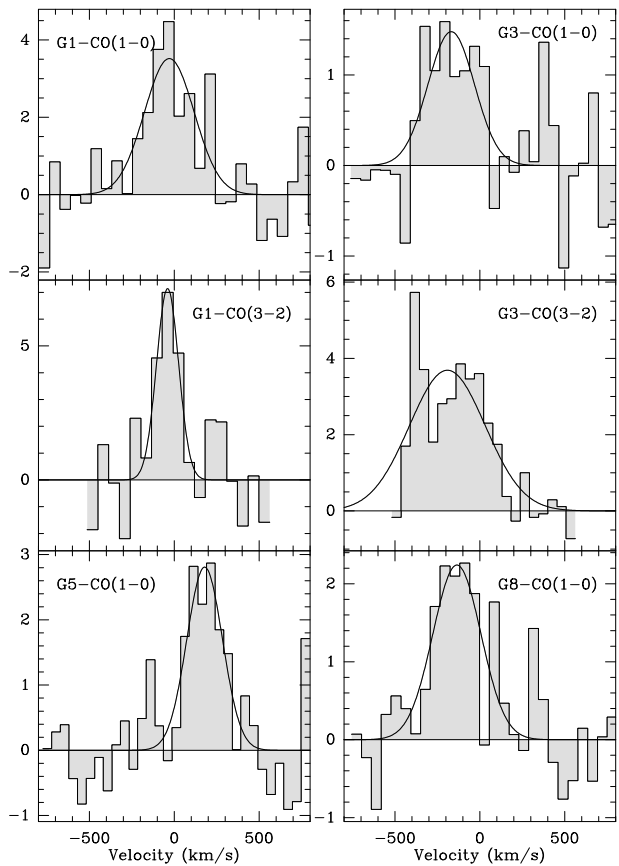}}
\caption{The CO spectra of the detected galaxies. The zero velocity scale
corresponds to the optically determined redshift,
listed in Table  \ref{tab:sample}. Sources
detected in CO(3--2) at 1mm wavelength are also shown. Some sources
were detected in CO(1--0) but
not in CO(3--2), as indicated in Table \ref{tab:lines}.
The vertical scale is T$_{mb}$ in mK. The spectrum of G4 is already presented in Paper I.}
\label{fig:spec1}
\end{figure}

\begin{figure}[!t]
\resizebox{8cm}{!}{\includegraphics[angle=0,width=8cm]{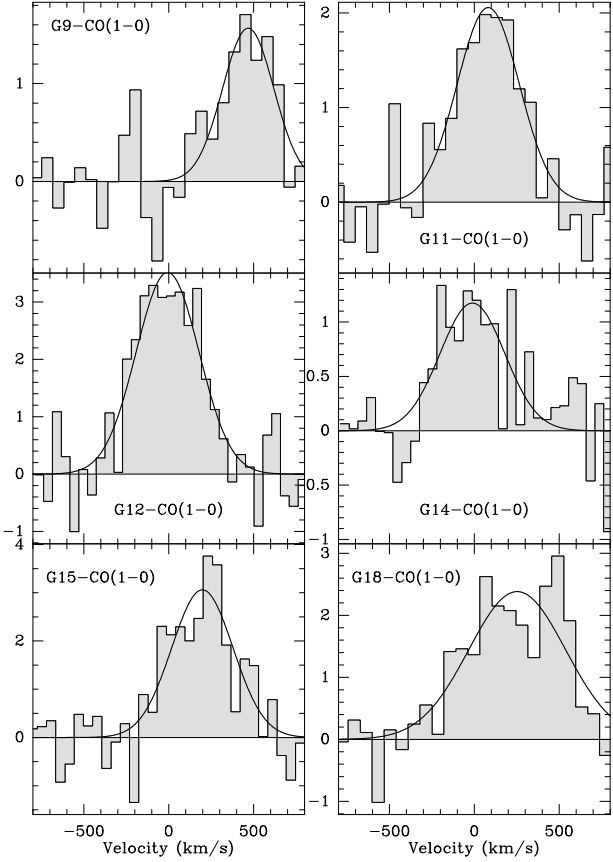}}
\caption{ Same as Fig. \ref{fig:spec1} for the following galaxies.}
\label{fig:spec2}
\end{figure}

\begin{figure}[!t]
\resizebox{8cm}{!}{\includegraphics[angle=0,width=8cm]{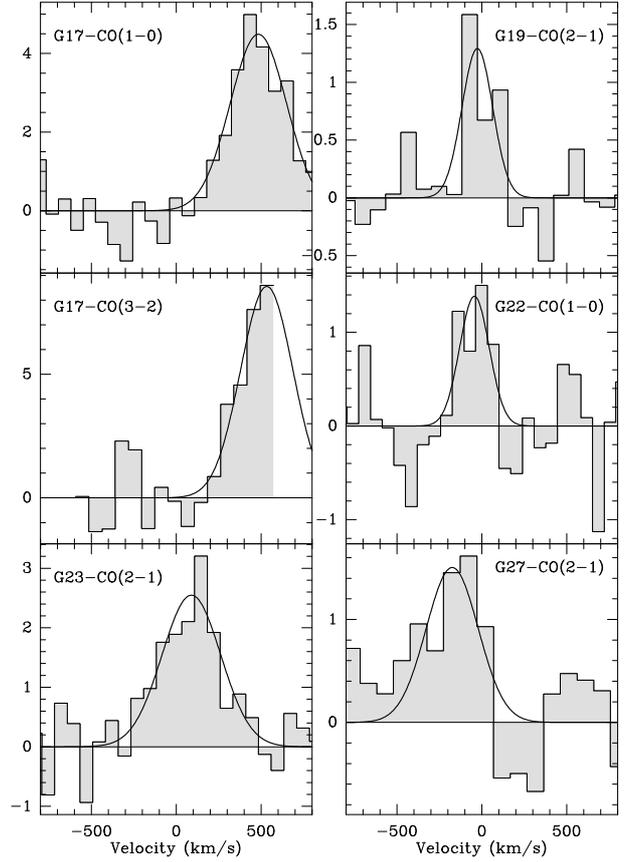}}
\caption{Same as Fig. \ref{fig:spec1}  for the remaining galaxies.}
\label{fig:spec3}
\end{figure}

The far-infrared fluxes F$_{\rm FIR}$ are 
 computed as 1.26 x 10$^{-14}$ (2.58 S$_{60}$+S$_{100}$) W m$^{-2}$ 
(Sanders \& Mirabel 1996). The far-infrared luminosity is then
 L$_{\rm FIR}$  = 4 $\pi$ D$_L^2$ CC F$_{\rm FIR}$, where 
D$_L$ is the luminosity distance, and
CC the color correction, CC=1.42 (e.g. Sanders \& Mirabel 1996).
The FIR-to-radio ratio 
$q$=log([F$_{\rm FIR}$/(3.75 10$^{12}$ Hz)]/[f$_\nu$(1.4 GHz)])
has been computed for sources where radio data were available;
 the radio fluxes are listed in Table \ref{tab:lines}.
Excluding the radio galaxies 3C48 and 3C345,
the average is $q$=2.3, typical for ULIRGs (Sanders \& Mirabel 1996). 
 The star formation rates of all sample galaxies are 
above 480 $\rm M_{\odot}yr^{-1}$, 
estimated from the infrared luminosity (e.g. Kennicutt 1998).

In this article, we adopt a standard flat cosmological model,
with $\Lambda$ = 0.73, and a Hubble constant of 71\,km\,s$^{-1}$\,Mpc$^{-1}$
(Hinshaw \etal\, 2009).

\begin{table}[h]
      \caption[]{Definition of the sample}
         \label{tab:sample}
            \begin{tabular}{l c c c c}
            \hline
            \noalign{\smallskip}
  G & Source   & RA(2000) & DEC(2000) & z \\
            \noalign{\smallskip}
            \hline
            \noalign{\smallskip}
G1& IRAS 00302+3625& 00:32:57.6& +36:41:56& 0.3023\\
G2& IRAS 08081+2611& 08:11:14.4& +26:02:17& 0.3850\\
G3& IRAS 10091+4704& 10:12:16.7& +46:49:43& 0.2460\\
G4& IRAS 11582+3020& 12:00:46.8& +30:04:15& 0.2230\\
G5& $^{a}$J12054771+1651085& 12:05:47.7& +16:51:08& 0.2170\\
G6& $^{b}$J1307006+233805& 13:07:00.6& +23:38:05& 0.2750\\
G7& $^{a}$J13301520+3346293& 13:30:15.2& +33:46:29& 0.3600\\
G8& IRAS 13352+6402& 13:36:50.7& +63:47:03& 0.2366\\
G9& IRAS 13379+3339& 13:40:14.4& +33:24:45& 0.2473\\
G10& IRAS 13447+2833& 13:47:05.5& +28:18:05& 0.2551\\
G11& IRAS 15298+6319& 15:30:41.1& +63:09:40& 0.2810\\
G12& IRAS 16300+1558& 16:32:21.4& +15:51:45& 0.2417\\
G13& [HB89] 1821+643& 18:21:57.3& +64:20:36& 0.2970\\
G14& IRAS 20551+2441& 20:57:19.7& +24:53:37& 0.2425\\
G15& IRAS 23113+0314& 23:13:54.3& +03:30:58& 0.3053\\
G16& IRAS 01506+2554 & 01:53:28.3 &+26:09:40 & 0.3264\\
G17& IRAS F02115+0226 & 02:14:10.3 &+02:40:00 & 0.4000\\
G18& IRAS 07449+3350 & 07:48:10.6 &+33:43:27 & 0.3560\\
G19& $^{b}$J0913454+405628 & 09:13:45.4 &+40:56:28 & 0.4420\\
G20& IRAS F10156+3705 & 10:18:34.5 &+36:49:52 & 0.4900\\
G21& IRAS 12514+1027 & 12:54:00.8 &+10:11:12 & 0.3000\\
G22&  [HB89] 1402+436 & 14:04:38.8 &+43:27:07 & 0.3233\\
G23& $^{c}$J145658.42+333710.1 & 14:56:58.4 &+33:37:10 & 0.4430\\
G24& IRAS 19104+8436 & 19:01:44.5 &+84:41:25 & 0.3544\\
G25& IRAS F00415-0737 & 00:44:05.6 &-07:21:13 & 0.3120\\
G26& $^{c}$J020412.43-005351.4 & 02:04:12.4 &-00:53:51 & 0.3343\\
G27& IRAS F00235+1024 & 00:26:06.5 &+10:41:32 & 0.5750\\
G28&  3C345 & 16:42:58.8 &+39:48:37 & 0.5928\\
G29&  3C48 & 01:37:41.3 & +33:09:35 & 0.3695\\
G30&  $^{c}$J140931.25+051131.2 & 14:09:31.2 & +05:11:31 & 0.2644\\ 
            \noalign{\smallskip}
            \hline
           \end{tabular}
[$^{a}$] 2MASX source;
[$^{b}$] 2MASSi source; 
[$^{c}$] SDSS source
\end{table}

\section{Observations}
\label{obs}

The observations were carried out with the IRAM 30m telescope
at Pico Veleta, Spain,
between January 2005 and  January  2006. Most of the 
galaxies, with redshifts between 0.2 and 0.39, could be observed
simultaneously
at 3mm in CO(1-0) and at 1mm in CO(3-2), except the two lowest
redshift sources (G4 and G5) where only observations in the 3mm band were possible.
In any case, the weather prevented us sometimes from taking
useful data at 1mm. For the highest redshift sources (G19, G20, G23, G27
and G28), only the CO(2-1) line was observed in the 2mm band.  

The SIS receivers were tuned in single sideband mode to the redshifted
frequencies of the various CO lines. The
observations were carried out in wobbler switching mode, with reference
positions offset by $3\arcmin$ in azimuth. We used the $1$~MHz
back-ends with an effective total bandwidth of $512$~MHz at $3$~mm
(providing a $\sim$ 1500 \kms\, range)
and the $4$~MHz filterbanks with an effective total bandwidth of
$1024$~MHz at $1$~mm.

We spent $2$--$4$ hours on each galaxy, resulting in a
relatively homogeneous noise level of $1$--$2$~mK per
$30$~km~s$^{-1}$ channel for all sources. The system temperatures
ranged between $120$ and $250$~K at $3$~mm, 
between $220$ and $300$~K at $2$~mm, 
and between $300$ and $500$~K at $1.2$~mm, in T$_A^*$.
The pointing was regularly checked on continuum sources and
yielded an accuracy of 3$''$ rms.  The temperature scale used is in main
beam temperature $T_{\rm mb}$. 
At 3mm, 2mm and 1mm, the telescope half-power beam
width is 27$''$, 17$''$ and 10$''$ respectively. 
The main-beam efficiencies are $\eta$$_{\rm mb}=T_{\rm
  A}^*/T_{\rm mb}$=0.85, 0.70 and 0.64, respectively, 
and $S/T_{\rm mb}$ = 4.8 Jy/K for all bands. 

Each spectrum was summed and reduced using linear baselines, and
then binned to $50-60$~km~s$^{-1}$ channels for the plots.

\begin{table*}
      \caption[]{Observed line parameters}
         \label{tab:lines}
            \begin{tabular}{l c c c c c c c c c c}
            \hline
            \noalign{\smallskip}
Galaxy &Line&$\nu_{\rm obs}$&S(CO)$^{a}$ & $V^{b}$&$\Delta V_{\rm FWHM}$&L'$_{\rm CO}$/10$^{10}$&S$_{60}$&S$_{100}^{c}$&
log L$_{\rm FIR}$ & F(1.4GHz)$^{d}$ \\
         &          & [GHz]             &  [Jy \kms]        & [\kms]      &[\kms]     & [K  \kms\, pc$^2$]&[Jy]&[Jy]& [\lsol] & [mJy]\\
            \noalign{\smallskip}
            \hline
            \noalign{\smallskip}
  G1  &  CO(1--0)  &  88.534  &   7.5  $ \pm$  1.4  &   -29.  $ \pm$   33.  &   398.  $ \pm$   98.  &     3.58 &   0.68 &   0.77 &  12.54 & \\
  G1  &  CO(3--2)  & 265.588  &   6.2  $ \pm$  2.0  &   -38.  $ \pm$   21.  &   148.  $ \pm$   62.  &     0.33 &   0.68 &   0.77 &  12.54 & \\
  G2  &  CO(1--0)  &  83.228 &  $<$  1.2  &  &  & $<$   0.9 &   0.67 &   1.13 &  12.84 & 4.09\\
  G2  &  CO(3--2)  & 249.672 &  $<$  3.2  &  &  & $<$   0.3 &   0.67 &   1.13 &  12.84 & 4.09\\
  G3  &  CO(1--0)  &  92.513  &   2.3  $ \pm$  0.5  &  -168.  $ \pm$   41.  &   314.  $ \pm$   66.  &     0.73 &   1.18 &   1.55 &  12.59 & 2.16\\
  G3  &  CO(3--2)  & 277.525  &    9.6  $ \pm$  0.9  &  -185.  $ \pm$   37.  &   498.  $ \pm$   47.  &     0.33 &   1.18 &   1.55 &  12.59 & 2.16\\
  G4  &  CO(1--0)  &  94.253  &   6.5  $ \pm$  0.7  &   248.  $ \pm$   24.  &   434.  $ \pm$   52.  &     1.67 &   1.23 &   1.52 &  12.50 & 3.09 \\
  G5  &  CO(1--0)  &  94.718  &   3.4  $ \pm$  0.6  &   178.  $ \pm$   20.  &   240.  $ \pm$   44.  &     0.83 &   1.36 &   1.54 &  12.51 & 25.5 \\
  G6  &  CO(1--0)  &  90.409 &  $<$  1.7  &  &  & $<$   0.7 &   0.72 &   0.69 &  12.44 & 3.0 \\
  G7  &  CO(1--0)  &  84.758 &  $<$  1.7  &  &  & $<$   1.2 &   1.18 &   1.20 &  12.94 & \\
  G7  &  CO(3--2)  & 254.262 &  $<$  4.5  &  &  & $<$   0.3 &   1.18 &   1.20 &  12.94 & \\
  G8  &  CO(1--0)  &  93.186  &   3.4  $ \pm$  0.7  &  -135.  $ \pm$   32.  &   301.  $ \pm$   82.  &     1.0 &   0.99 &   1.43 &  12.49 & 5.96\\
  G9  &  CO(1--0)  &  92.439  &   2.1  $ \pm$  0.5  &   469.  $ \pm$   30.  &   290.  $ \pm$   72.  &     0.67 &   0.94 &   1.10 &  12.48 & 3.85\\
  G10  &  CO(1--0)  &  91.850 &  $<$  1.7  &  &  & $<$   0.6 &   0.82 &   1.04 &  12.46 & 7.44\\
  G11  &  CO(1--0)  &  89.985  &   4.5  $ \pm$  0.6  &    85.  $ \pm$   30.  &   428.  $ \pm$   66.  &     1.86 &   0.84 &   1.15 &  12.58 & 2.77\\
  G12  &  CO(1--0)  &  92.811  &   7.7  $ \pm$  0.8  &    -5.  $ \pm$   21.  &   430.  $ \pm$   44.  &     2.33 &   1.48 &   1.99 &  12.68 & 5.67\\
  G13  &  CO(1--0)  &  88.875 &  $<$  1.7  &  &  & $<$   0.8 &   1.24 &   2.13 &  12.84 & 95.\\
  G14  &  CO(1--0)  &  92.811  &   2.4  $ \pm$  0.7  &   -10.  $ \pm$   70.  &   420.  $ \pm$  140.  &     0.73 &   1.06 &   1.43 &  12.53 & \\
  G15  &  CO(1--0)  &  88.330  &   6.9  $ \pm$  1.1  &   200.  $ \pm$   34.  &   433.  $ \pm$   73.  &     3.38 &   1.22 &   1.27 &  12.79 & \\
  G16  &  CO(1--0)  &  86.932 &  $<$  1.2  &  &  & $<$   0.7 &   0.58 &   1.14 &  12.63 & 40.3\\
  G17  &  CO(1--0)  &  82.337  &   8.3  $ \pm$  0.9  &   478.  $ \pm$   19.  &   363.  $ \pm$   46.  &     7.14 &   0.32 &   0.64 &  12.59 & \\
  G17  &  CO(3--2)  & 246.997  &   12.9  $ \pm$  2.0  &   500.  $ \pm$   22.  &   303.  $ \pm$   56.  &     1.24 &   0.32 &   0.64 &  12.59 & \\
  G18  &  CO(1--0)  &  85.008  &   7.6  $ \pm$  0.7  &   253.  $ \pm$   32.  &   651.  $ \pm$   64.  &     5.18 &   0.61 &   1.22 &  12.75 & 2.3 \\
  G18  &  CO(3--2)  & 255.012 &  $<$  4.5  &  &  & $<$   0.3 &   0.61 &   1.22 &  12.75 & 2.3\\
  G19  &  CO(2--1)  & 159.874  &   1.5  $ \pm$  0.4  &   -28.  $ \pm$   30.  &   225.  $ \pm$   63.  &     0.40 &   0.53 &  $<$0.44 &  12.78 & 6.87\\
  G20  &  CO(2--1)  & 154.723 &  $<$  1.4  &  &  & $<$   0.5 &   0.23 &   0.57 &  12.70 & 8.82 \\
  G21  &  CO(1--0)  &  88.670 &  $<$  1.7  &  &  & $<$   0.8 &   0.71 &   0.76 &  12.54 & 7.78\\
  G21  &  CO(3--2)  & 265.997 &  $<$  4.5  &  &  & $<$   0.2 &   0.71 &   0.76 &  12.54 &7.78 \\
  G22  &  CO(1--0)  &  87.129  &   1.5  $ \pm$  0.3  &   -41.  $ \pm$   26.  &   196.  $ \pm$   44.  &     0.82 &   0.62 &   0.99 &  12.62 & 1.59\\
  G22  &  CO(3--2)  & 261.373 &  $<$  2.3  &  &  & $<$   0.1 &   0.62 &   0.99 &  12.62 & 1.59\\
  G23  &  CO(2--1)  & 159.763  &   4.7  $ \pm$  0.7  &    91.  $ \pm$   25.  &   370.  $ \pm$   60.  &     1.27 &   0.23 &   0.62 &  12.61 & 1.76 \\
  G24  &  CO(1--0)  &  85.134 &  $<$  0.8  &  &  & $<$   0.6 &   0.52 &   1.83 &  12.80 & \\
  G24  &  CO(3--2)  & 255.388 &  $<$  4.5  &  &  & $<$   0.3 &   0.52 &   1.83 &  12.80 & \\
  G25  &  CO(1--0)  &  87.859 &  $<$  1.2  &  &  & $<$   0.6 &   0.40 &   1.06 &  12.49 & \\
  G25  &  CO(3--2)  & 263.564 &  $<$  9.0  &  &  & $<$   0.5 &   0.40 &   1.06 &  12.49 & \\
  G26  &  CO(1--0)  &  86.410 &  $<$  1.2  &  &  & $<$   0.7 &   0.46 &   0.83 &  12.54 & 1.36\\
  G26  &  CO(3--2)  & 259.217 &  $<$  4.5  &  &  & $<$   0.3 &   0.46 &   0.83 &  12.54 & 1.36\\
  G27  &  CO(2--1)  & 146.373  &   2.1  $ \pm$  0.6  &  -126.  $ \pm$   34.  &   231.  $ \pm$   90.  &     0.95 &   0.43 &  $<$0.94 &  13.11 & 2.7 \\
  G28  &  CO(2--1)  & 144.719 &  $<$  1.9  &  &  & $<$   0.9 &   0.60 &   1.26 &  13.28 & 7000\\
  G29$^{e}$  &  CO(1--0)  &  84.170  &   1.9  $ \pm$  0.3 &    -6.  $ \pm$   10.  &   270.  $ \pm$   20.  &     1.42 &   0.74 &   0.83 &  12.78 & 16000\\
  G30$^{f}$  &  CO(1--0)  &  91.123  &   3.8  $ \pm$  0.4 &     185.  $ \pm$   20.  &   270.  $ \pm$   30.  &     1.40 &   1.45 &   1.82 &  12.75 & 5.49\\
            \noalign{\smallskip}
            \hline
           \end{tabular}
\begin{list}{}{}
\item[] Quoted errors are statistical errors from Gaussian
            fits. The systematic calibration uncertainty is 10\%. 
\item[$^{a}$] The upper limits are at 3$\sigma$ with an assumed $\Delta$V = 300 \kms.
$^{b}$ The velocity is relative to the optical redshift given in Table \ref{tab:sample}.  
\item[$^{c}$] The 60 and 100\mic\, fluxes are from NED (http://nedwww.ipac.caltech.edu/)
\item[$^{d}$] From the FIRST  catalog (http://sundog.stsci.edu/). Errors are typically 0.14 mJy
\item[$^{e}$] From Wink \etal\, (1997).  $^{f}$ From Solomon \etal\, (1997).
\end{list}
\end{table*}

\section{Results}
\label{res}

\subsection{CO detection in z=0.2-0.6 ULIRGs}

All spectra for CO detections are displayed in 
Figures \ref{fig:spec1}, \ref{fig:spec2} and \ref{fig:spec3}
(except G4 reported in paper I). The non-detections are
reported in Table \ref{tab:lines}.
Integrated upper limits are computed at 3$\sigma$,
assuming a common line-width of 300\kms, and getting
the rms of the signal over 300\kms. Lines are assumed detected 
when the integrated signal is larger than 3$\sigma$. Gaussian fits then yielded the
central velocities, velocity FWHMs and integrated
fluxes listed in Table~\ref{tab:lines}. 

As already noticed in Sect. \ref{sample},  very few objects 
were previously detected in CO  in this redshift range. We include
in our analysis, and in Table \ref{tab:lines},
two additional ULIRGs (G29 and G30) that satisfy our sample criteria.
In the discussion, we also added the two galaxies on the outskirts of
the cluster Cl 0024+16 (Geach \etal\, 2009); they are not ULIRGs,
but it is interesting to compare star formation efficiencies for
all CO-detected objects in this redshift range.

The detection rate of 60\% in our sample, down to a sensitivity
limit of $\sim$ 1.5 Jy\kms, must be considered a lower
limit. Indeed, the available velocity range of the receivers (about 1500\kms)
could have missed some sources if the optical redshift was not 
accurate enough. Some galaxies show a significant velocity offset  (e.g. G17)
as can be seen in Table \ref{tab:lines} and the figures. Some of the profiles may have a double-horn
shape as G18, but most do not, given our spectral resolution and sensitivity.
  The line-widths detected are compatible with massive galaxies
at random inclinations. Their average is $\Delta$V$_{\rm FWHM}$= 348 \kms,
very similar to the value for local ULIRGs of 302\kms\, (Solomon \etal\, 1997).
In comparison, the submillimeter galaxies have much broader widths,
655 \kms\, on average (Greve 2005).
 Given the angular distance of the sources (average value 1000 Mpc), our beam
subtends between 50 and 100kpc, and all galaxies can be considered unresolved,
at least as far as their molecular component is concerned.

\subsection{CO luminosity and \hh\, mass}
\label{SFE}

To derive the total \hh\, mass, we first compute the CO luminosity
through integrating the CO intensity over the velocity profile.

The CO luminosity for a high-z source is given by $$L'_{CO} =
23.5 I_{CO} \Omega_B {{D_L^2}\over {(1+z)^3}} \hskip6pt \rm{K\hskip3pt
  km \hskip3pt s^{-1}\hskip3pt pc^2}$$
where  $I_{CO}$ is the intensity in K \kms, $\Omega_B$ is the area of
the main beam in square arcseconds and $D_L$  is the luminosity
distance in Mpc.  We assume here that the sources are unresolved,
in our beam of typically 50-100kpc.
We then compute H$_2$ masses using
M$_{\rm H_2} = \alpha$ L'$_{\rm CO}$, 
with $\alpha=0.8$ M$_\odot$ (K \kms\, pc$^2$)$^{-1}$,
for ULIRGs. The molecular gas masses are
listed in Table~\ref{tab:h2}. Although it could be advocated
that a different conversion factor should apply to some of the galaxies,
we always refer to the M(\hh) mass, directly proportional
to CO luminosity, for the sake of comparison.
For those galaxies where we observed two CO transitions,
we used the CO luminosity of the lower transition to calculate \hh\, masses.
In our sample,
most galaxies have CO(1-0) data, except three galaxies,
G19, G23 and G27, which have been detected in CO(2-1). We assume that the
brightness temperatures are similar in the two lines, as expected 
for an optically thick, and thermally excited medium. The problem
is more severe for high-z objects, where the CO excitation is not
well-known.
The average CO luminosity for the 18 galaxies detected
is L'$_{\rm CO}$ = 2. 10$^{10}$ K \kms\, pc$^2$, corresponding to an 
average \hh\, mass of  1.6 10$^{10}$ \msol.

The star formation efficiency (SFE), also listed in Table~\ref{tab:h2},
is defined as  L$_{\rm FIR}$/M(\hh) in \lsol/\msol. Since the SFR
is related to the FIR luminosity as
SFR= L$_{\rm FIR}$ /(5.8 10$^9$L$_{\odot}$) (e.g. Kennicutt 1998),
the gas consumption time-scale can be derived by 
$\tau$ = 5.8/SFE  Gyr.

\begin{table}[h]
      \caption[]{Molecular gas mass and star formation efficiency}
         \label{tab:h2}
            \begin{tabular}{l c c c c c}
            \hline
            \noalign{\smallskip}
  G & M(\hh)   & SFE  & T$_d$ & M$_d$ & Type\\
&  10$^9$ M$_\odot$ &  \lsol/\msol\, &  K   &  10$^8$ M$_\odot$ &  \\
            \noalign{\smallskip}
            \hline
            \noalign{\smallskip}
   G1 &   28.6 &    120. &  51.6  &     0.8 & \\
   G2 & $<$7.5 & $>$910. &  47.0  &     2.6 & \\
   G3 &    5.8 &    667. &  46.5  &     1.6 & L, Int\\
   G4 &   13.4 &    238. &  46.7  &     1.2 & L\\
   G5 &    6.7 &    483. &  48.1  &     1.1 & L, Int\\
   G6 & $<$5.3 & $>$521. &  54.4  &     0.5 & S1,Q\\
   G7 & $<$9.4 & $>$928. &  56.5  &     1.2 & \\
   G8 &    7.9 &    391. &  44.5  &     1.6 & Pair\\
   G9 &    5.3 &    565. &  48.7  &     0.9 & S2, Int\\
  G10 & $<$4.6 & $>$637. &  47.4  &     1.0 & \\
  G11 &   14.9 &    255. &  46.9  &     1.4 & Int\\
  G12 &   18.6 &    255. &  46.0  &     2.0 & L, Int\\
  G13 & $<$6.3 &$>$1110. &  43.8  &     4.0 & S1,Q\\
  G14 &    5.8 &    584. &  45.7  &     1.5 & \\
  G15 &   27.0 &    227. &  53.5  &     1.1 & \\
  G16 & $<$5.3 & $>$802. &  42.7  &     2.9 & \\
  G17 &   57.1 &     67. &  44.9  &     2.0 & \\
  G18 &   41.4 &    134. &  43.5  &     3.4 & \\
  G19 &    3.2 &   1875. &$>$60.0  &  $<$0.5 & Q2, Int\\
  G20 & $<$3.8 &$>$1305. &  44.6  &     2.8 & \\
  G21 & $<$6.4 & $>$542. &  52.9  &     0.7 & S2, Int\\
  G22 &    6.6 &    626. &  45.9  &     1.8 &S1,Q, Int \\
  G23 &   10.2 &    400. &  42.1  &     3.3 & \\
  G24 & $<$4.5 &$>$1379. &  36.6  &    11.9 &S1,Q \\
  G25 & $<$4.9 & $>$631. &  38.5  &     4.0 & \\
  G26 & $<$5.6 & $>$615. &  44.2  &     1.9 & \\
  G27 &    7.6 &   1699. &$>$49.0  & $<$4.2 & \\
  G28 & $<$7.3 &$>$2615. &  50.3  &     5.3 &Q \\
  G29 &   11.4 &    523. &  54.5  &     1.0 &Q \\
  G30 &   11.2 &    497. &  48.0  &     1.9 & S2\\
            \noalign{\smallskip}
            \hline
           \end{tabular}
\\L: LINER, S1, S2: Seyfert 1 \& 2, Q: QSO, Int: interaction
\\ M(\hh) and SFE are defined in Sec. \ref{SFE}
\end{table}

\subsection{Molecular gas excitation}
\label{excit}

Three sources have been detected in both the
CO(1-0) and CO(3-2) lines, and three have upper limits,
as listed in Table \ref{tab:COex}.
For the data points, we took the peak flux S$_\nu$ of the lines,
since the CO(3-2) and CO(1-0) have sometimes different measured linewidths,
which could be due partly to the noise.  The corresponding ratio
between the peak brightness temperatures are also displayed in Table
\ref{tab:COex}, to compare more easily with the predictions of the model.
 
The peak flux ratio between the two lines S$_{32}$/S$_{10}$,
and equivalently the peak brightness temperature ratio, 
is a good indicator of the average density of the emitting medium, since 
density is the main factor determining the excitation.  Another factor is the kinetic temperature,
which could be linked to the dust temperature (e.g. Weiss \etal\, 2003).  In Table~\ref{tab:h2},
we have computed the dust temperature deduced from the far-infrared fluxes, 
 assuming $\kappa_\nu\propto\nu^{\beta}$, where
$\kappa_\nu$ is the mass opacity of the dust at frequency $\nu$, and
$\beta$ = 1.5. The average dust temperature for our sample
 is $46\pm5$~K. This is comparable to recent results for starburst galaxies, which have dust
temperatures $\approx40$~K (e.g. Sanders \& Mirabel 1996,  Elbaz \etal\, 2010).

 If the gas is predominantly heated by collisions with the dust, the gas temperature
is expected to be lower than the dust temperature, at low density (e.g. Spitzer 1978). Alternatively, if the gas
is heated directly from UV photons near star forming regions, or by shocks
due to turbulence or perturbed dynamics, then
the gas could be at much higher kinetic temperatures.
Since we observe very low excitation temperatures, we consider it unlikely that,
in average over the beam, the hot molecular gas is dominating the emission.
 We then assume a gas kinetic temperature lower than the dust
temperature, in the following modeling.

\begin{figure*}
\centering
\includegraphics[angle=0,width=17cm]{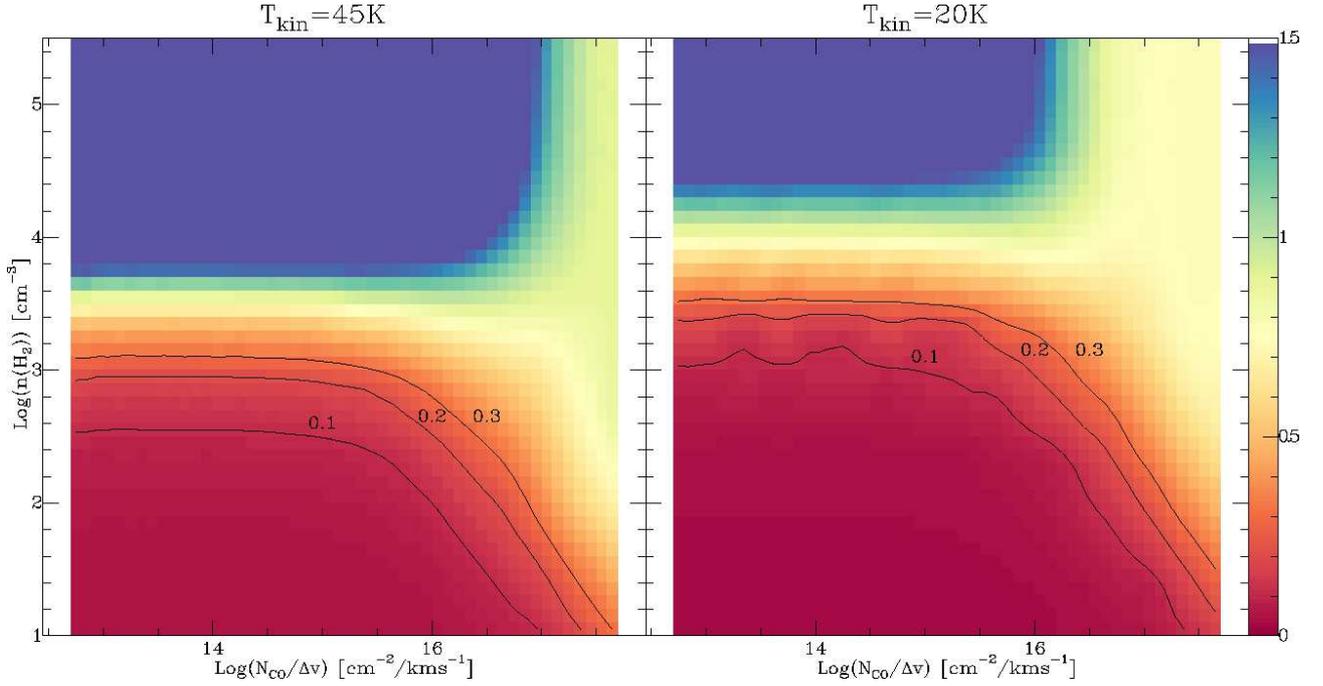}
\caption{Peak T$_{\rm b}$ ratio between the CO(3-2) and CO(1-0) lines
versus the \hh\, density, and the CO column density per unit velocity
width (N$_{\rm CO}/\Delta V$) for two values of the  kinetic temperature:
T$_k$ = 45K, the dust temperature (left), andT$_k$ = 20K (right).
The black contours are underlining the values
obtained in the data. The predictions come from the LVG hypothesis
in the Radex code.} 
\label{fig:lvg}
\end{figure*}

 Using the Radex code (van der Tak \etal\, 2007), we have computed
the predicted main beam temperature ratio between the 
CO(3-2) and CO(1-0) lines, for several kinetic temperatures
and as a function of \hh\, densities and CO column densities.
Figure \ref{fig:lvg} shows these predictions for T$_k$=45 and 20K. 
The black contours delineate the range of observed values.
 In Table  \ref{tab:COex} we list the derived values for the n(\hh) densities,
for two values of the kinetic temperatures (45 and 20K), and for a fixed
column density per velocity width.

We adopted a column density of N(CO)/$\Delta V$
of 7$\times$10$^{16}$ cm$^{-2}$/(\kms), which assumes that all CO  
lines are optically thick. This number is at the right order of magnitude
given the high molecular gas masses derived in Table \ref{tab:lines}. 
For M(\hh) = 3 10$^{10}$ M$_\odot$, a typical CO abundance of 
CO/\hh\, = 10$^{-4}$, and a linewidth of 300 \kms,
this column density corresponds to a homogenous 
disk of 3 kpc in size. Either the emitting CO gas is more concentrated,
as in nuclear starbursts, i.e. the CO column density would be higher,
or the gas extent is larger than 3kpc, in which case we would have to take the
clumping factor into acount. In any case, it is likely that the CO lines are optically
thick. 

The excitation of the CO gas appears quite low, implying
a low average \hh\, density in our galaxies.
Comparing to the different excitation patterns observed in other high-z
starburst galaxies (Weiss \etal\, 2007), our galaxies are among the lowest 
excitation, comparable to the Milky Way or even lower. 

For the estimations, we have assumed that our sources are unresolved.
Increasing the size of the
molecular disks, here supposed to be point like (wrt to the beam sizes) to 
for example $>$6'' would increase the T$_b$ ratios by $>$50\%, 
and this would raise
the required densities by, at most, a similar factor according to 
Fig. \ref{fig:lvg}. 

It should be kept in mind that error bars are large on the observed ratios.
The average ratio is lower than 1.8$\pm$0.6, taking into account the
upper limits. However the conclusion of a rather low average \hh\, density
is  rather robust with respect to the error bars, since the predicted flux ratio
is increasing very quickly with density in the model. We note that even 
with the extreme hypothesis of optically thin gas, the observations
are only compatible with n(\hh) $<$ 3 10$^3$ cm$^{-3}$, since the 
predicted ratio increases even more with density than in the thick case. 
 It is also possible that some of the gas has a kinetic temperature much
higher than the dust temperature, but then the derived \hh\, density is
even lower.

\subsection{Variation with redshift}

\begin{figure}[ht!]
\centering
\includegraphics[angle=0,width=8cm]{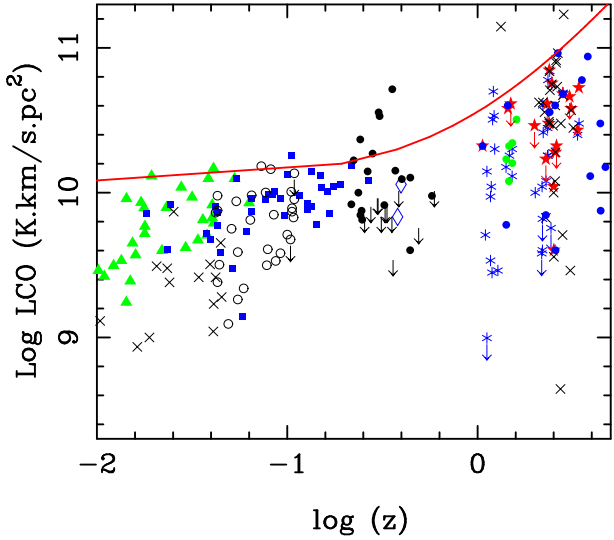}
\caption{Measured CO luminosities, corrected for
amplification when known, but not for gas excitation,
as a function of redshift.  We compare 
our points (full black circles, and arrows as upper limits) 
with a compilation of high-z molecular gas surveys, and local ones:
green triangles are from Gao \& Solomon (2004),
blue squares from Solomon \etal\, (1997),
open circles from Chung \etal\, (2009),
blue diamonds from Geach \etal\, (2009),
black crosses from Iono \etal\, (2009),
red stars, from Greve \etal\, (2005),
 green full circles from Daddi \etal\, (2010),
blue asterisks from Genzel \etal\, (2010), and  blue full circles 
from Solomon \& vanden Bout (2005). 
For illustration purposes only,
the red curve is the power law for $\Omega_\hh$/$\Omega_\hi$
proposed by Obreschkow \& Rawlings (2009).}
\label{fig:CO-z}
\end{figure}

Does the molecular gas content of galaxies evolve with redshift?
It is interesting to compare the CO luminosity
of our sample with the wealth of data reported in
the literature.  Figure \ref{fig:CO-z} shows all CO measurements
as a function of redshift. This figure reveals that indeed our sample (full black circles)
is filling  in the CO redshift desert, although not completely.  The rise of the 
CO luminosity that is observed at high redshift (z $>$1) in fact begins
as soon as z=0.2-0.3.

This variation is meaningful in the sense that only the brightest objects
have been selected here. Most of the variation with z comes from the fact
that there are no extremely luminous objects at z$<$ 0.2. In itself, it is already an 
interesting evolution, that has been discussed in previous works at
high redshift (i.e. Solomon \& vanden Bout 2005, Tacconi \etal\, 2010).
The present work extends this variation in the intermediate redshift range,
and suggests that the increase in gas content with z might begin as soon as z=0.3.
The possibility of undiscovered large CO luminosity objects locally
is not high, given the good correlation between CO and FIR luminosity.
These objects should have been discovered as ULIRGs.

To interpret this evolution, caveats have to be kept in mind. At high
redshift, at least for some of the sources,
the CO luminosity could be over-estimated by poorly known
amplification factors due to lensing. 
The luminosities have been corrected for amplification,
when known, but these uncertainties contribute to  the large scatter.
This is not the case for
the sample from Daddi \etal\, (2010, green circles) or the sample
from Genzel \etal\, (2010, blue asterisks).
Another uncertainty comes from the CO gas excitation. The CO luminosities
 of high redshift ULIRGs come from the measured high-J lines, and the low-J
lines are often not known. They could underestimate the \hh\, mass,
since the subthermal excitation is likely 
to reduce their luminosity with respect to the local objects, observed in CO(1-0).
  We think, however, that the steep rise at 0.2 $<$ z $<$ 0.6 of the most
luminous  galaxies discussed in
this paper does not suffer from these caveats (they are detected in
majority in CO(1-0) and are not lensed).
 It is interesting to note that the trend seen in Fig \ref{fig:CO-z} is essentially dominated 
by three galaxies, G1, G15 and G18, which are particularly strong in CO-emission.
No such extreme  L'$_{\rm CO}$ has  been found in the local universe.
 Two of these galaxies (G1 and G18) have been shown in Sect. \ref{excit}
to be subthermally excited, and are likely extended starbursts with relatively
low efficiency, as presented by Daddi \etal\, (2008). 
In paper I, we also derived an extended gas disk for
G4 with the Plateau de Bure observations.
Table \ref{tab:h2} confirms that G1 and G18 have among 
the lowest SFE of the sample. It is conceivable
that the conversion factor between L'$_{\rm CO}$ and M(\hh) could be
higher in these sources, leading to higher gas mass.

\subsection{Correlation between FIR and CO luminosities}

\begin{figure}[ht!]
\centering
\includegraphics[angle=0,width=8cm]{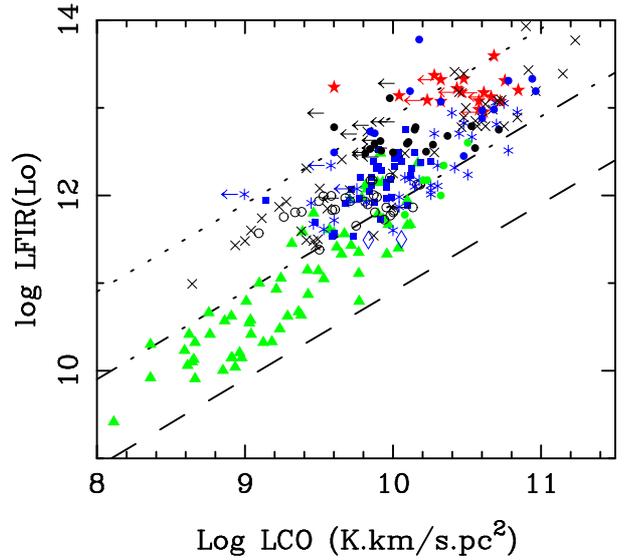}
\caption{Correlation between FIR and CO luminosities, for 
our sample (full black circles, and arrows for upper limits) 
and the other points from the literature (same symbols
as in Fig \ref{fig:CO-z}). The 3 lines are for L$_{\rm FIR}$/M(\hh)=10,
100 and 1000 \lsol/\msol\, from bottom to top, assuming a conversion factor
$\alpha$ = 0.8  M$_\odot$ (K \kms\, pc$^2$)$^{-1}$. 
The three lines correspond to gas depletion time-scales of 580, 58 and 5.8 Myr
respectively.}
\label{fig:FIR-CO}
\end{figure}

Figure \ref{fig:FIR-CO} shows the well studied correlation
between FIR and CO luminosities (e.g. Young \& Scoville 1991). 
The correlation  is non-linear, with the ultra-luminous objects displaying a 
higher FIR  luminosity (i.e. star formation) for the amount of gas present,
as implied by the 
lines plotted in the Figure.  Our sample galaxies fit perfectly in this
picture, being all above the curve  L$_{\rm FIR}$/M(\hh)=100 \lsol/\msol\,
(corresponding to a consumption time-scale of $\tau$ =58 Myr).
One of the two galaxies detected above  L$_{\rm FIR}$/M(\hh)=1000 \lsol/\msol\,
(G19) has indications of nuclear activity (see Type in Table \ref{tab:h2}), but
the second (G27) has none. We note that they are two of the three galaxies
detected in CO(2-1), and not in CO(1-0); we have derived their \hh\, masses
with the assumption of equal CO luminosity between these two first lines.
 If their gas was sub-thermally excited,
 their \hh\, mass could then be slightly under-estimated.

Assuming a dust temperature T$_d$ and the observed 100~\mic\,
flux S$_{100}$, we can derive the dust mass as
$$
\begin{array}{lcl}
 {\rm M}_d & = & 4.8\times10^{-11}\, {{S_{\nu o}\,D_{\rm Mpc}^{\,2}}
 \over {(1+z)\kappa_{\nu r}\,B_{\nu r}(T_d)}}\ \msol \\
             &  =&  5(1+z)^{-(4+\beta)}\,S_{100\mu}\,D_{\rm Mpc}^{\,2}\,
  \left\{\exp(144(1+z)/T_d) - 1 \right\}\ \msol\, \\
\end{array}
$$ 
where $S_{\nu o}$ is the observed FIR flux measured in Jy, $D_{\rm Mpc}^{\,2}$ is the luminosity
distance in Mpc, $B_{\nu r}$ is the Planck function at the rest
frequency $\nu r = \nu o (1+z)$, and we use a mass
opacity coefficient of $25$~cm$^{2}$~g$^{-1}$ at rest-frame 100~\mic, 
(Hildebrand 1983,  Dunne \etal\, 2000, Draine 2003), with a frequency dependence of $\beta$=1.5.
Estimated dust masses are displayed in Table \ref{tab:h2}.
 If we adopt the low conversion factor of $\alpha$ = 0.8  M$_\odot$ (K \kms\, pc$^2$)$^{-1}$,
the average gas-to-dust mass ratio is 96 for all the detected galaxies.
 The gas to dust mass ratio could increase to up to 550 if the standard (MW)
conversion factor is used. For local ULIRGs, this number is 
around 100 (Solomon \etal\, 1997), and 700 for normal galaxies
(Wiklind \etal\, 1995), when calculated from IRAS fluxes.

We should note that the dust temperature has been measured 
from 60 and 100\mic\, fluxes, which correspond to (1+z) higher 
frequencies in the rest frame of the galaxies.  Therefore we are not sensitive 
to the very cold dust ($\sim$ 10K).

\begin{table}[h]
\centering
      \caption[]{CO gas excitation}
         \label{tab:COex}
            \begin{tabular}{l c c c c}
            \hline
            \noalign{\smallskip}
  G &  S$_{32}$/S$_{10}$ &    T$_{\rm b32}$/T$_{\rm b10}$         & n(\hh) &   n(\hh) \\
    &    [peak]                &     [peak]        & [cm$^{-3}$] &    [cm$^{-3}$] \\
    &                               &                       &             45K         &        20K   \\
            \noalign{\smallskip}
            \hline
            \noalign{\smallskip}
   G1 &   2.2 $\pm$ 1.1       &   0.25 $\pm$ 0.1   &  63. & 200. \\
   G3  &  2.6  $\pm$ 0.8    &     0.29 $\pm$ 0.09  &  79. & 250. \\
  G17  & 1.9   $\pm$ 0.5   &    0.21 $\pm$ 0.05 &   46. &  140. \\
  G18  &  $<$ 1.3             &   $<$   0.14   &   $<$20. &    $<$63. \\
 G22  &  $<$ 1.0               &  $<$  0.11   &   $<$12.  & $<$40.  \\
             \noalign{\smallskip}
            \hline
           \end{tabular}
\\ n(\hh), for T$_k$=45K and 20K, and N$_{\rm CO}/\Delta V$=7$\times$10$^{16}$ cm$^{-2}$/(\kms)
\end{table}

\subsection{Activity of the galaxies}

\begin{figure*}[ht!]
\centering
\includegraphics[angle=0,width=17cm]{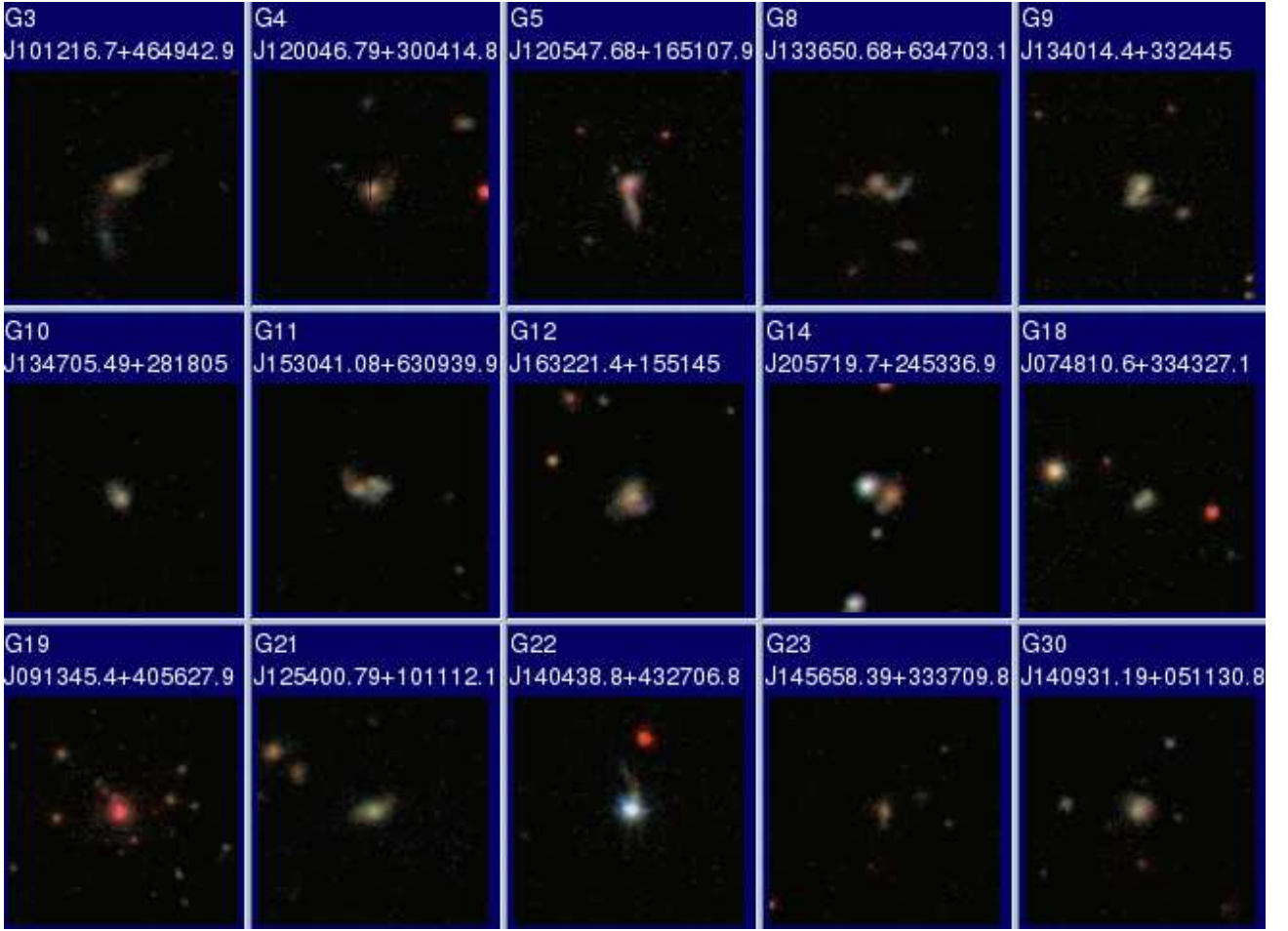}
\caption{Optical color images from the Sloan Digital Sky Survey 
(SDSS, http://www.sdss.org/) of 15 of our sources. They are all detections,
except G10 and G21. Each panel is 50$''$x50$''$ in size, and is centered
on the galaxy coordinates of Table \ref{tab:sample}. 
}
\label{fig:sdss}
\end{figure*}

We have made a census of the different activities occuring
in our sample galaxies.  The last column of Table \ref{tab:h2} indicates 
nuclear AGN activity and/or perturbed morphology.
 These have been derived from the SDSS images, some of which are
shown in Figure \ref{fig:sdss}.
  We note that among the 12 non-detections, there are only 4 
active objects, all being Seyfert 1 or quasars, while among the 18 detected
ones, there are 12 active objects, and most of the time they are LINERs, Seyfert 2,
and show signs of galaxy interactions and mergers.

\subsection{Star formation efficiency}

\begin{figure}[ht!]
\centering
\includegraphics[angle=0,width=8cm]{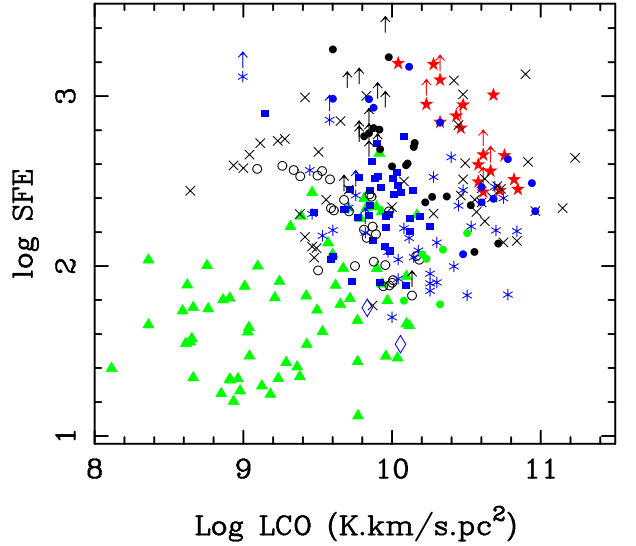}
\caption{Star formation efficiency SFE=L$_{\rm FIR}$/M(\hh),
versus CO luminosity, assuming the same CO-to-\hh\, conversion factor
$\alpha$ = 0.8  M$_\odot$ (K \kms\, pc$^2$)$^{-1}$. All symbols are as defined
in Fig \ref{fig:CO-z}. }
\label{fig:SFE-CO}
\end{figure}

\begin{figure}[ht!]
\centering
\includegraphics[angle=0,width=8cm]{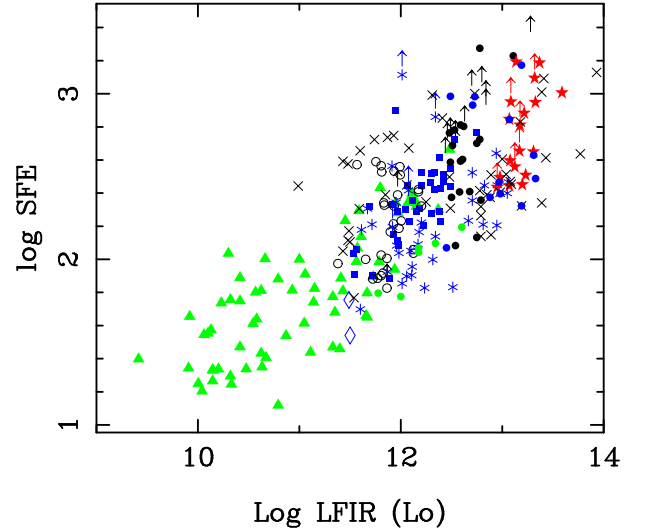}
\caption{Same as Figure  \ref{fig:SFE-CO}, but versus L$_{\rm FIR}$. }
\label{fig:SFE-FIR}
\end{figure}

\begin{figure}[ht!]
\centering
\includegraphics[angle=0,width=8cm]{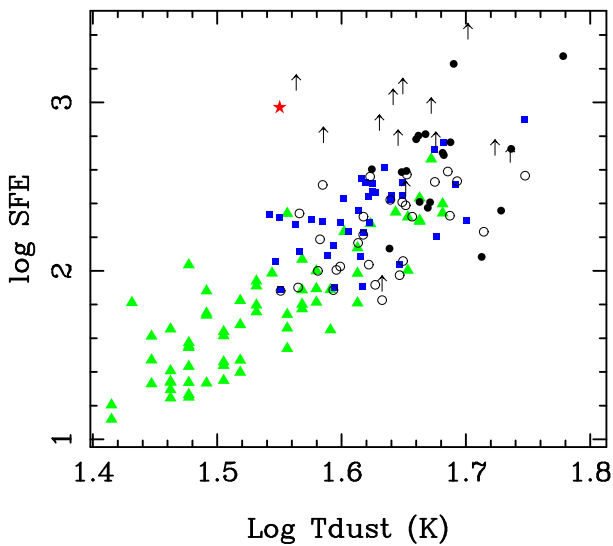}
\caption{Same as Figure  \ref{fig:SFE-CO}, but versus T$_d$, for the
sources where it could be defined. For the submillimeter galaxies, the red star
corresponds to the averaged SFE, with the mean dust temperature of 35.5 found
by Kovacs \etal\, (2010).}
\label{fig:SFE-Td}
\end{figure}

In the following we adopt the definition of SFE = L$_{\rm FIR}$/M(\hh),
with a constant CO-to-\hh\, conversion factor.
The average SFE in our sample is 555 \lsol/\msol,
3 times higher than that of the local ULIRGs (170 \lsol/\msol).
It should be kept in mind that some
galaxies could have  a different conversion factor, and this uncertainty
affects our conclusions. 
Also, it is possible that the local SFR tracers evolve with time,
and that, for a given SFR, different amounts of gas are consumed
in star formation, if the IMF (Initial Mass Function)
 is different for earlier and younger galaxies. However, no strong evidence has been
found until now for a significantly changing IMF, and the star formation laws are remarkably
constant over redshift, as discussed by Genzel \etal\, (2010) and Daddi \etal\, (2010).
We plot the SFE versus L'$_{\rm CO}$ in 
Figure \ref{fig:SFE-CO}, versus L$_{\rm FIR}$ in 
Figure \ref{fig:SFE-FIR}, and versus the dust temperature in
Figure \ref{fig:SFE-Td}.

What is obvious in all these figures is that galaxies of our sample are among
the most efficient forming stars, and G19 and G27 even
lie above the starbursts at high redshift. They are not among
the most gas-rich, according to the CO luminosity. They could be experiencing a burst 
due to galaxy interactions. This is the case for G19 (Figure \ref{fig:sdss}).
No image is available for G27
(which lies outside of the footprint of the SDSS).
Note that, as expected, there is a much larger correlation between SFE and L$_{\rm FIR}$,
than with  L'$_{\rm CO}$. The presence of large amounts of gas is not a sufficient condition
to trigger a starburst, and another hidden factor is the extent of the spatial distribution of the molecular gas.

\begin{figure}[ht!]
\centering
\includegraphics[angle=-90,width=8cm]{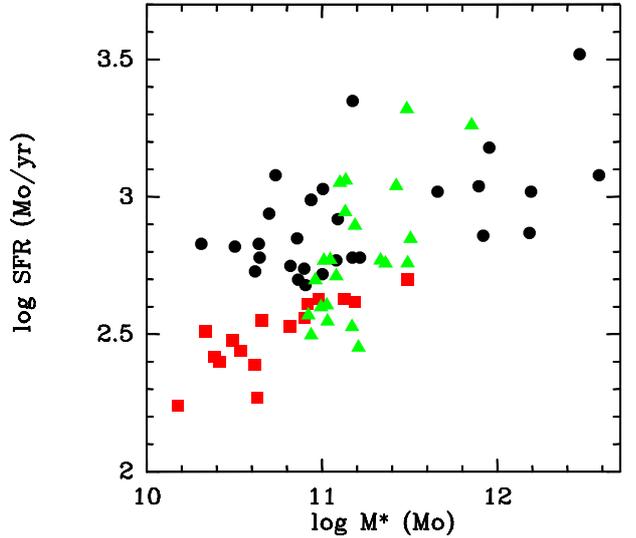}
\caption{The star formation rate (SFR) obtained from the far infrared luminosity,
versus the stellar mass of galaxies in our sample (full black circles), compared
to  the sample of Da Cunha et al (2010, full red squares) and Fiolet et al (2009, full green triangles).}
\label{fig:SFR-Ms}
\end{figure}

To estimate the relative gas fractions of our sample galaxies, an estimation 
of their stellar and/or dynamical mass is required. We have tried to estimate the stellar mass from 
optical and near-infrared luminosities, taken from the literature, 
mainly the SDSS and 2MASS catalogs. The multi-wavelength luminosities were
K-corrected according to the colors (e.g. Chilingarian \etal\, 2010), and stellar masses
estimated according also to the colors (Bell et al 2003). Stellar masses were found between
10$^{10}$ and 10$^{12}$ M$_\odot$ or somewhat higher in the case of quasars.
Figure \ref{fig:SFR-Ms} displays the star formation rate, derived from the infrared luminosity,
versus the stellar mass, in comparison to the ULIRG samples of Da Cunha et al (2010) 
and Fiolet et al (2009). Our sample points follow the general trend,
with the bias of strong SFR, due to our selection on L(FIR). The gas fractions derived
from these stellar masses show large variations, with 3 objects around 1\%,
but most of them are between 10\% and 65\%.
Another estimation is the dynamical mass, derived from the observed CO linewidths.
This can only be a rough estimate, since neither the inclinations of the galaxies,
nor the extent of their CO emission, are known. Adopting a typical radius of 3kpc,
the dynamical masses are in the range of 10$^{11}$ M$_\odot$, and the derived gas fraction are in
general a few percent, with a large scatter, up to 60\%.
Note that the derived stellar masses are on average larger than the dynamical masses;
this is due to the radius we have selected (3kpc) to estimate the dynamical masses. For the same
velocity width, the dynamical mass scales as the radius.
 Firm conclusions cannot be drawn until CO maps are available. Indeed, it is possible
that the sample has a wide range of radial extents and, consequently, conversion factors.

\begin{figure}[ht!]
\centering
\includegraphics[angle=0,width=8cm]{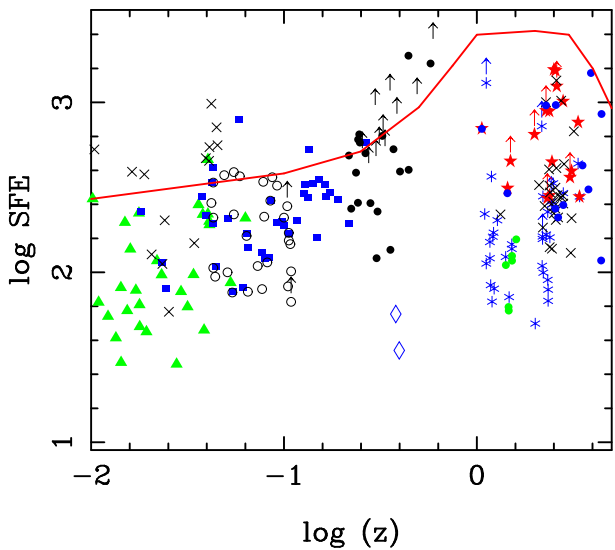}
\caption{Same as Figure  \ref{fig:SFE-CO}, but versus redshift. The red curve
is a schematic line summarizing the cosmic star formation history, 
from the compilation by Hopkins \& Beacom (2006), complemented with the 
GRB data by Kistler \etal\, (2009) and the optical data from Bouwens \etal\, (2008).
The red curve is logarithmic and
only indicative of relative variations of the star formation rate per cubic Mpc
as a function of redshift, and can be translated vertically.}
\label{fig:SFE-z}
\end{figure}

Finally, there is a correlation between the dust temperature and the SFE in
Figure \ref{fig:SFE-Td}: it is conceivable that 
a concentrated starburst heats more efficiently the dust
around it. However, the correlation becomes more scattered at higher redshifts,
where the largest efficiencies occur. It is possible that our estimation
of the dust temperature is not as accurate for these more redshifted objects.
It is interesting to note the evolution of SFE with redshift, where our
two most efficient starbursts are clearly noticeable, as shown in Figure \ref{fig:SFE-z}.
 We qualitatively compare this evolution with the cosmic star formation
history, as compiled by Hopkins \& Beacom (2006), from 
different works in the literature, and complemented at very high redshift
by the gamma-ray burst (GRB) data of Kistler \etal\, (2009)
and the optical data (Lyman-Break Galaxies, LBG) from Bouwens \etal\, (2008).
The SFE logarithmic variations should be a combination between variations of the SFR
and of the gas fraction in galaxies. It is interesting to superpose the observed 
logarithmic curve of these SFR variations 
 to have an indication of the relative role of the various parameters.
 The schematic curve in log reproduces the relative variations,
whatever the vertical units, and can be arbitrarily translated vertically.
Our points correspond to the most drastic change in this curve, and 
our following study at 0.6 $<$ z $<$ 1 should  give more insight
in this epoch.

\section{Discussion and conclusions}
\label{disc}
We have presented our search for CO-line emission in a sample of 30 
ULIRGs, selected between 0.2 $<$  z $<$ 0.6 to fill the gap or ``CO redshift desert'' between
z=0.1 and 1. We intend to cover the second part 0.6 $<$  z $<$ 1.0  in a following work.
Our detection rate is $\sim$ 60\%.  We find that some of the galaxies
possess large amounts of molecular gas, much larger than local ULIRGs.
  Considering the evolution with cosmic time, it appears that the huge amounts
of gas, common at high redshift, begin to disappear at z$\sim$ 0.3.
This drop in gas content is reminiscent of the drop in the star formation 
history of the universe, which may imply
that the change in star formation is due to a 
change in gas content.  There are good reasons to think that galaxies are more
gas rich at high redshift, and also that their gaseous medium is denser.
The sizes of galaxies are predicted to vary as (1+z)$^{-1}$, and the implied
higher gas pressure could increase the \hh/\hi\, ratio. Following semi-analytical 
simulations, Obreschkow \etal\, (2009) followed the \hh/\hi\, ratio statistically
over 30 million galaxies, and its cosmic decline has been modelled as
$\Omega_\hh$/$\Omega_\hi \propto (1+z)^{1.6}$ by
Obreschkow \& Rawlings (2009). This law appears to reproduce grossly
the decline in the maximum L'$_{\rm CO}$ with time, as shown in Figure \ref{fig:CO-z}.
 Some galaxies of our sample, however, lie significantly above this envelope.
 The increase with z in the \hh\, content of galaxies might occur 
already at lower z than this model predicts.

Five of our galaxies were observed in both CO(3-2) and CO(1-0) lines, 
allowing an estimation of the excitation temperature. They appear all very low,
similar to  what is observed in the Milky Way, or more normal galaxies, but also
some local ULIRGs (Radford \etal\, 1991).
These ULIRGs could be similar to those discovered by Daddi \etal\, (2008)
at redshift z$\sim$1.5. If a galactic conversion factor was adopted for these galaxies,
as suggested by Daddi \etal\, (2010),
their \hh\, mass would be even higher,
and they would stand out even more in the cosmic \hh\, abundance.

 We emphasize that the choice of the conversion factor is crucial for
the interpretation of the results. We have adopted the ULIRG value proposed
by Solomon \etal\, (1997), and derive large SFE and short consumption time-scales
for the gas. These SFE values would be lower if a higher, i.e. the Galactic, conversion factor is used.
However, for consistency with previous studies we prefer to assume only a single value for the conversion factor.
 The latter could vary with the extent of
the molecular component. We do not yet have spatial information on the CO emission and
future interferometer observations are required to constrain the conversion factor further.

We have compared the star formation history and the redshift evolution of the 
SFE. It is expected that the latter evolves as a combination of the SFR
and gas fraction evolution.
It is likely that the star formation decline between z=1 and z=0 is
partly due to the declining star forming efficiency. For galaxies of our sample,
the star formation efficiency (SFE) appears very high, in comparison to
the most active starbursts at different redshifts.
This supports a high contribution of the SFE to the star formation
variations with redshift, although
we are observing an increase of efficiency of the most extreme objects.
It is possible to compare the observed time gradients in 
the cosmic star formation rate, and those in the extreme SFE 
in Figure \ref{fig:SFE-z}.  We observe a significant 
 gradient in SFE, but however less steep than in the star formation
history. The latter requires  also a strong variation in gas content.

The very efficient star forming objects (ULIRGs) 
dominate the star formation at high redshift  (z $<$ 1.5, Lefloc'h \etal\, 2005), 
and less extreme objects (LIRGs) dominate later on (Caputi \etal\, 2006),
which might explain the strong decline in efficiency between z=1 and 0.
It appears that the range of redshift studied here is just
 where the most massive objects continue
to form stars with unprecedented efficiency,
before the sudden drop due to star formation quenching
(e.g. Springel \etal\, 2005).

\begin{acknowledgements}
  We warmly thank the referee for his/her constructive comments
and suggestions. 
The IRAM staff is gratefully acknowledged for their
help in the data acquisition.
We have made use of the NASA/IPAC Extragalactic Database (NED).
\end{acknowledgements}

\end{document}